\documentclass [12pt,a4paper,notitlepage,dvips]{article}
\usepackage{dcolumn,bm,amsmath}
\usepackage{graphicx}        
\setkeys{Gin}{draft=false}   
\usepackage[sort&compress]{natbib} 
\bibpunct[, ]{[}{]}{,}{n}{,}{,}    
\newcommand{\eref}[1]{(\ref{#1})}
\newcommand{\Eref}[1]{Eq.\,(\ref{#1})}
\newcommand{\Tref}[1]{Table\,\ref{#1}}

\begin{document}
\title{Precision calculations of atoms with few valence electrons}
\author{ M.\,G.\,Kozlov\\
\textit{Petersburg Nuclear Physics Institute, Gatchina,
188300, Russia}\\
E-mail:\texttt{mgk@MF1309.spb.edu}}
\date{\today}
\maketitle

\begin{abstract}
We discuss the possibility of using pair-equations for the construction of
the effective Hamiltonian $H_{\rm eff}$ for valence electrons of an atom.
The low-energy part of atomic spectrum is found by solving the eigenvalue
problem for $H_{\rm eff}$. In this way it is possible to account
efficiently for the valence-valence and core-valence correlations. We
tested this method on a toy model of a four-electron atom with the core
$1s^2$. The spectrum obtained with $H_{\rm eff}$ for two valence electrons
is in a perfect agreement with the full configuration interaction for all
four electrons.
\end{abstract}

\section*{Introduction}

The reliable and accurate \textit{ab initio} methods for atomic
calculations are necessary not only for atomic physics and quantum
chemistry, but also for application of atomic physics to the investigation
of the fundamental interactions. For example precision calculations of Cs
were necessary to test electroweak theory on the level of radiative
corrections at low energy (see \cite{Hag02} and references therein). Recent
search for the possible variation of the fine structure constant $\alpha$
\cite{WMF01} in space-time also required large-scale atomic calculations
\cite{DFK02}.

At present there are several widely used methods for calculations of
many-electron atoms. For atoms with one valence electron above the core the
many-body perturbation theory (MBPT) appears to be very effective
\cite{SDJ98}. For the atoms with more than one valence electrons the
accuracy of the conventional MBPT techniques is not satisfactory mainly
because of the poor convergence of the MBPT for the valence-valence
correlations. Because of that non-perturbative methods are usually
preferable \cite{Pal92,PFG96,Eli96,EK96,MET00}. However, the core-valence
correlations can still be effectively treated with MBPT. Because of that it
was suggested to combine MBPT for the core-valence correlations with the
configuration interaction (CI) for valence-valence correlations within the
CI+MBPT method \cite{DFK96b}.

In this method CI calculation for valence electrons is done with the
effective Hamiltonian $H_{\rm eff}$. This Hamiltonian accounts for core
polarization and screening of the two-electron interaction between
valence electrons. It may be formed within the Brillouin-Wigner variant of
MBPT. In practice most of the calculations were done within the second
order MBPT \cite{KP97,DJ98,PKR00b,KPJ01,SJ02}. The operator $H_{\rm eff}$ is
defined for valence electrons only. Therefore, CI+MBPT method is equivalent
to the multi-reference MBPT with the model space coinciding with the whole
valence space. The use of the Brillouin-Wigner variant of MBPT allows
to avoid problems with intruder states, but leads to the energy-dependent
effective Hamiltonian~\cite{HW00,KP99tr}.

CI+MBPT method has several important limitations:
\begin{enumerate}
\item
The number of valence electrons should not be large. The accuracy of the
CI method rapidly decreases when this number exceeds 3 or 4.
\item
Only the lower part of the valence spectrum can be presented accurately.
For the energies above the excitation energy of the core the effective
Hamiltonian has poles and results of the CI+MBPT method may become
unreliable~\cite{PRK99a}.
\item
The accuracy of the calculations is limited because the effective
Hamiltonian does not account for the higher order core-valence
correlations. With the second order effective Hamiltonian the typical
accuracy for the transition frequencies is of the order of a percent.
\end{enumerate}
The first two limitations are intrinsic to the method. In order to address
the third one it is necessary to go beyond the second order MBPT. One
obvious way to do this is the order-by-order approach. However, it is known
that for atoms with one valence electron the third order calculations are
often even less accurate than the second order ones. Besides, for atoms
with more than one valence electron the third order is already too
complicated for practical use.

The arguments given above lead us to one of the variants of the all-order
methods. The coupled-cluster method is one of the most widely
used~\cite{Pal92}. It gives the nonlinear system of equations for
cluster amplitudes. In order to truncate this system it is often restricted
to SD excitations. The linearized version of the coupled-cluster equations
in the SD-approximation are also known as pair equations, or SD-method
\cite{BJLS89}. It was used for the calculations of the atoms with one
valence electron and proved to be significantly more accurate than the
second order MBPT~\cite{SJ00a}. So, we suggest to use CI+SD method
instead of CI+MBPT for the calculations of the atoms with few valence
electrons.


In order to use the SD-method for the effective Hamiltonian, we developed
the Brillouin-Wigner variant of these equations and added equations for the
two-electron amplitudes, which were absent in the one-electron atoms. That
allowed us to form the effective Hamiltonian for valence electrons in
SD-approximation.

As a test system for the CI+SD method we study a toy model of a 4-electron
atom with two valence electrons above the core $1s^2$. We use a very short
basis set, for which the 4-electron full CI is possible. That allows to
test the method without addressing such questions as saturation of the
basis set and the role of the Breit interaction, which we do not include
here. This model was recently considered in~\cite{Koz03} and appeared
useful for the study of the higher orders in the residual core-valence
interaction.

\section{Effective Hamiltonian}

It is well known that if we split the total many-electron Hilbert space
in two subspaces $P$ and $Q$, $P+Q=1$, the full Schr\"{o}dinger equation
can be equivalently rewritten in a following form:
\begin{eqnarray}
     &&\Psi = P \Psi+Q \Psi
     \equiv \Phi+\chi,
\label{Heff6}\\
     &&\left(PHP+ \Sigma(E)\right)
     \Phi =  E\Phi,
\label{Heff12}
\end{eqnarray}
where
\begin{eqnarray}
     &&\Sigma(E) = (PHQ)\ {R}_Q(E)\ (QHP),
\label{Heff13}\\
     &&R_Q(E) = (E-QHQ)^{-1}
\label{Heff10}
\end{eqnarray}
Eq.~\eref{Heff12} shows, that the effective
Hamiltonian in the subspace $P$ has the form:
\begin{eqnarray}
      H_{\rm eff}(E) \equiv PHP + \Sigma(E).
\label{Heff1}
\end{eqnarray}
Note that this $P,Q$ expansion is formally equivalent, for example, to the
exclusion of the small component from the Dirac equation. For practical
purposes the particular choice of the subspace $P$ is very important. For
atoms with few valence electrons it is natural to associate this subspace
with the valence one. In other words $P$ subspace corresponds to the
frozen-core approximation $P=P_{\rm FC}$. This means that core-core and
core-valence correlations are accounted for on the stage of formation of
the operator $H_{\rm eff}$, while \textit{all} valence correlations are
treated explicitly when solving~\Eref{Heff12}.

Of course, for many-electron atoms the exact form of the operator $R_Q(E)$
is not known and some approximations should be used. In principle, MBPT
gives consistent procedure for finding an order-by-order expansion of this
operator in powers of the residual two-electron interaction. In the lowest
(second) order $R_Q(E)$ is substituted by $R_Q^0(E) = (E-QH_0Q)^{-1}$,
where $H_0$ is a one-electron operator. This leads to the second order
effective Hamiltonian for valence electrons, which was used in calculations
\cite{KP97,DJ98,PKR00b,KPJ01}. The complete form of the third order
expansion for $H_{\rm eff}$ is still unknown and is probably to
complicated for practical use.

The most simple and compact variant of the MBPT expansion corresponds to
the choice, when $H_0$ is equal to the Dirac-Fock operator of the core
$H^{\rm DF}_{\rm core}$. At least for atoms with one or two valence
electrons this may be also an optimal choice~\cite{Koz03}. For more valence
electrons one may want to include, at least partly, the field of the
valence electrons in $H_0$. This leads to many new terms in the MPBT
expansion \cite{DFK96b}, but gives better initial approximation. In our
calculations we use the Dirac-Fock code \cite{BDT77}, which allows for the
flexible choice of the self-consistent field.

In the end of this section we want to mention that sometimes the whole
wave function \eref{Heff6} is needed. Then we can recover it from the
solution of \eref{Heff12} with the help of relation:
\begin{eqnarray}
     \Psi = \Phi+\chi =\left(1+ R_Q(E)\ (QHP)\right)\Phi.
\label{Heff11}
\end{eqnarray}
We again can use MBPT expansion for $R_Q(E)$ here. If we need $\chi$ to
calculate some atomic amplitudes, we can use \Eref{Heff11} to define
corresponding effective operator in the subspace $P$. For the effective
operators one can use the random phase approximation with additional
two-electron terms instead of the order-by-order MBPT~\cite{DKPF98}. This
technique was used in calculations of $E1$-transitions \cite{PKR00b},
polarizabilities \cite{KP99}, and hyperfine constants~\cite{PRK99a}.

\section{SD-method for the effective Hamiltonian}

As we mentioned above, the accuracy of the CI+MBPT method is limited
because of the neglect of the higher order corrections to $H_{\rm eff}$.
The SD-method allows to sum to all orders the MBPT terms, which correspond
to the one and two holes in the core in all intermediate cross-sections of
the Goldstone diagrams. Technically this leads to the substitution of the
one- and two-electron matrix elements in $H_{\rm eff}$ with corresponding
SD-amplitudes.

The explicit form of the SD-equation depends on the choice of the
one-electron operator $H_0$, which is used for the initial approximation.
The simplest form corresponds to $H_0=H^{\rm DF}_{\rm core}$, which we
assume below. We expect that in analogy with the CI+MBPT, this variant of
the CI+SD method should be good at least for one- and two-electron
atoms.

\begin{figure}[t]
\begin{center}
\includegraphics[scale=1.0]{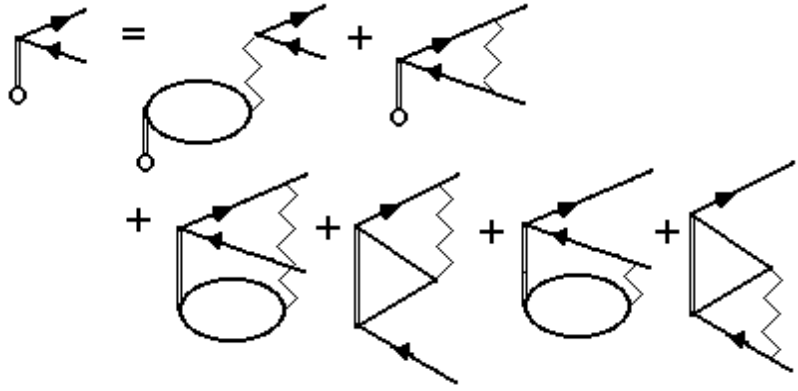}
\end{center}
\caption{Graphic form of the SD-equation for the one-electron core
amplitude. Wavy line corresponds to Coulomb interaction, double line
denotes two-electron SD-amplitude, and double line with the circle denotes
one-electron SD-amplitude.}
\label{fig1}
\end{figure}

\begin{figure}[hb]
\begin{center}
\includegraphics[scale=1.0]{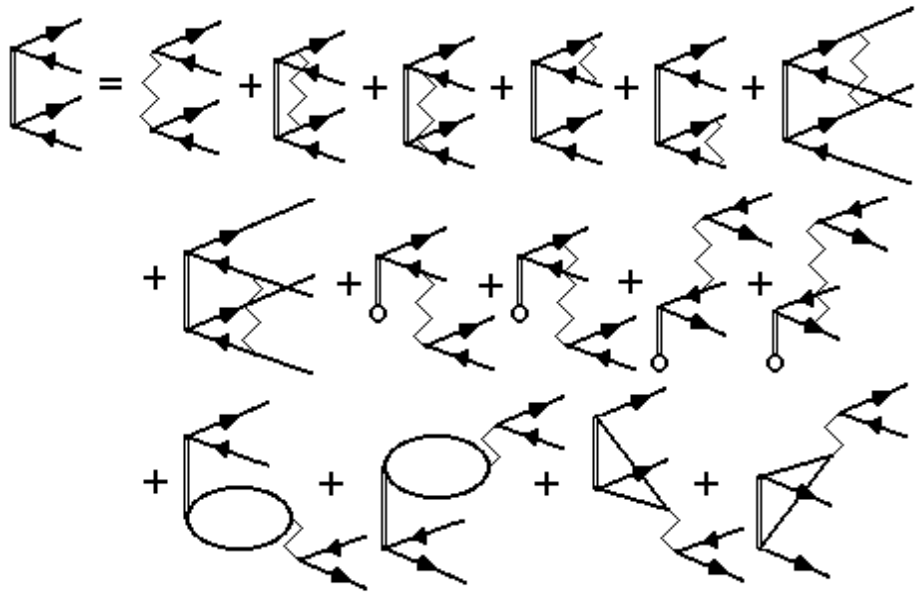}
\end{center}
\caption{Graphic form of the SD-equation for the two-electron core
amplitude.}
\label{fig2}
\end{figure}

The system of equations for SD-amplitudes splits in three parts. The first
subsystem includes only amplitudes for core excitations. Therefore, these
amplitudes do not depend on core-valence and valence amplitudes and should
be solved first \cite{BJLS89}. The graphic representation of this subsystem
is shown in Fig.~\ref{fig1} and~\ref{fig2}. Note that all equations are
presented in a form, suitable for iterative solution. At first iteration we
put all SD-amplitudes on the right-hand-sides to zero. That leaves the
single non-zero term in the equation for the two-electron SD-amplitude
Fig.~\ref{fig2}. As a result we get non-zero two-electron amplitude, but
the one-electron SD-amplitude is still zero. On the next iteration both
right-hand-sides in Fig.~\ref{fig1} and~\ref{fig2} are non-zero.

\begin{figure}[htb]
\begin{center}
\includegraphics[scale=1.0]{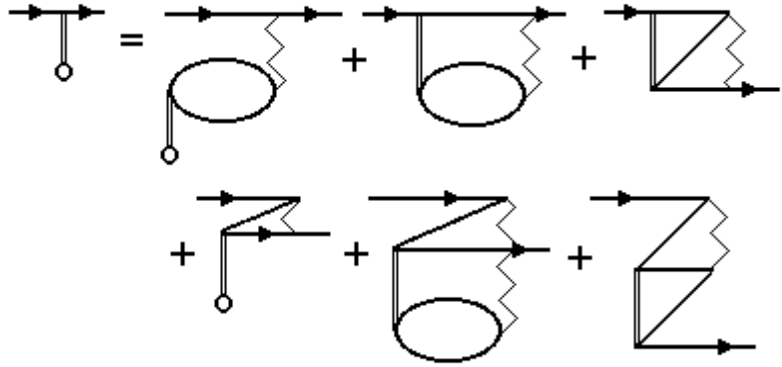}
\end{center}
\caption{SD-equation for the one-electron valence amplitude.}
\label{fig3}
\end{figure}

\begin{figure}[htb]
\begin{center}
\includegraphics[scale=1.0]{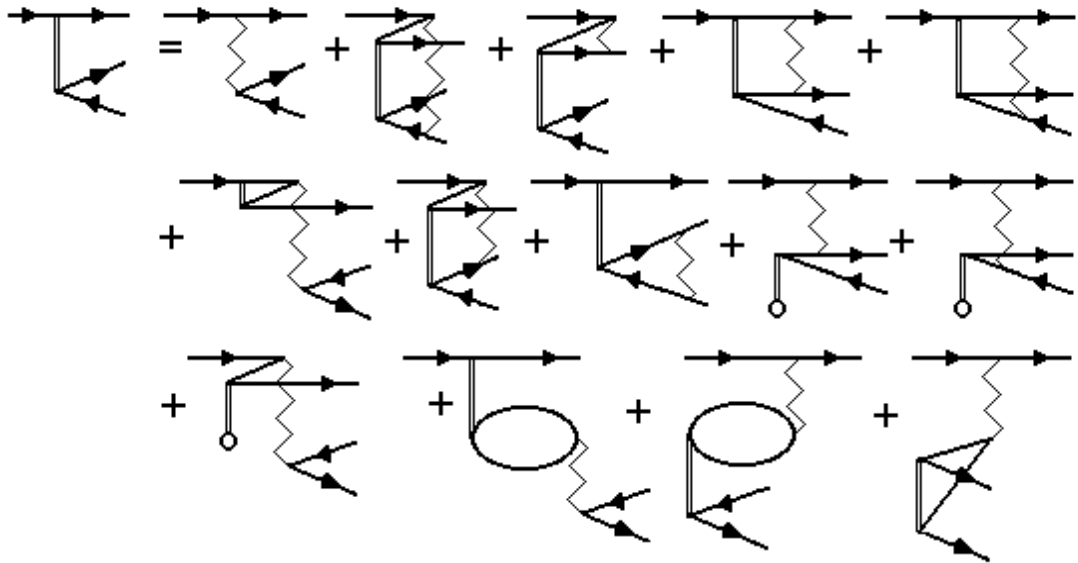}
\end{center}
\caption{SD-equation for the two-electron core-valence amplitude.}
\label{fig4}
\end{figure}

On the next step the one electron valence amplitudes and two-electron
core-valence amplitudes should be found from the system of equations shown
in Fig.~\ref{fig3} and~\ref{fig4}. It is seen that they depend on each
other and on the core amplitudes, which were found on the previous step.
This system again can be solved iteratively. Iteration processes on the
first and the second steps converge rather rapidly because the energy
denominators for \textit{all} diagrams are large, because they include the
excitation energy of the core. The latter grows with the number of
valence electrons, so we can expect faster convergence for the atoms with
more valence electrons.

\begin{figure}[htb]
\begin{center}
\includegraphics[scale=1.0]{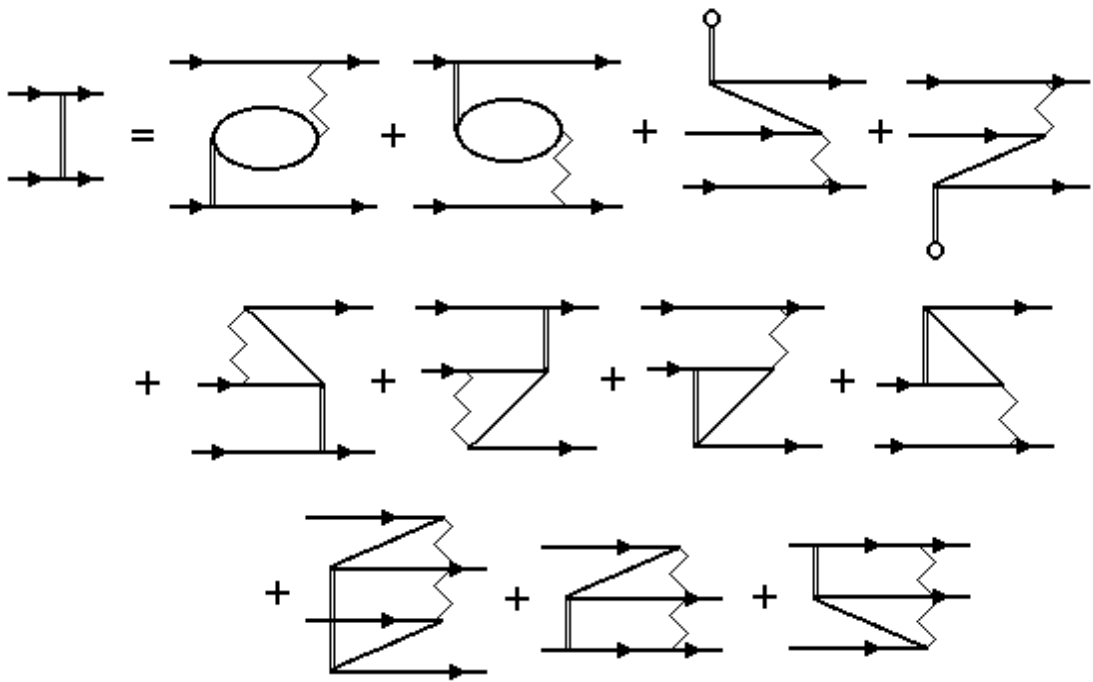}
\end{center}
\caption{Diagramatic expression for the two-electron valence amplitude.}
\label{fig5}
\end{figure}

The one-electron valence SD-amplitudes, which are found from the equations
on Fig.~\ref{fig3} can be already used for the construction of $H_{\rm
eff}$. However, on this step we still do not have two-electron valence
amplitudes. These can be found by calculating diagrams from
Fig.~\ref{fig5}. Corresponding diagrams depend only on the amplitudes,
which are already found on previous steps, so we can calculate two-electron
valence SD-amplitudes in one run. Therefore, the third step is rather
simple and this is the only new step, which was not used in calculations of
one-electron atoms~\cite{BJLS89}. It means that the SD-method is easily
generalized for the many-electron case.

There are several questions, which we did not address above. First is the
energy dependence of the SD-amplitudes. That can be accounted for in a same
way, as it was done in CI+MBPT method. Second, we prefer to have Hermitian
effective Hamiltonian, while the valence SD-amplitudes as given by
Fig.~\ref{fig3} and~\ref{fig5} are non-symmetric. Thus, we need to
symmetrize them somehow. These questions will be discussed elsewhere in
detail. Here we prefer to give an example of the application of the CI+SD
method to a model system, which can be solved `exactly' and the accuracy of
the CI+SD approximation can be therefore well controlled and compared to
that of the CI+MBPT method.

\section{Toy model}

Here we consider the atom with $Z=6$ and 4 electrons in a very short basis
set of 14 orbitals $1-4s_{1/2}$, $2-4p_j$, and $3,4d_j$. Because of such a
short basis set, we can not compare our results with the physical spectrum
of C~III. Instead, we can do the full CI for all 4 electrons and use it as
an `exact' solution. Alternatively, we can consider the same atom as a
two-electron system with the core $[1s^2]$. In this case we can use both
CI+MBPT and SI+SD methods and compare their results with the known `exact'
solution.

\begin{figure}[htb]
\begin{center}
\includegraphics[scale=1.0]{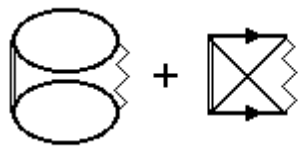}
\end{center}
\caption{Core correlation energy in the SD-approximation.}
\label{fig6}
\end{figure}

As an even simpler test systems we can also consider the same system with 2
and 3 electrons. The core equations Fig.~\ref{fig1} and~\ref{fig2} are the
same for all three cases. After they are solved, we can immediately find
the correlation correction to the core energy by calculating two diagrams
from Fig.~\ref{fig6}. The same correction can be found with the help of
the full CI for two electrons. Note that SD-method for the two electron
system is exact, so both result should coincide. In this way we can
effectively check the system of equations for core amplitudes.

The iterative procedure for the core SD-equations converges in three steps
and the core correlation energy $\delta E_{\rm core}$ in atomic units is
given by:
\begin{center}
\begin{tabular}{ccc}
Iteration & $\delta E_{\rm core}$ & Difference \\
    1     &  $-0.006051$   &$-0.006051$ \\
    2     &  $-0.006278$   &$-0.000228$ \\
    3     &  $-0.006282$   &$-0.000004$ \\
 Full CI  &  $-0.006280$   &$+0.000002$ \\
\end{tabular}
\end{center}

The difference $2\cdot 10^{-6}$~a.u. between the full CI and SD results is
two orders of magnitude smaller than the error of the second order MBPT,
which corresponds to the first iteration and is probably due to the
numerical accuracy (we store radial integrals in the CI code \cite{KT87} as
\texttt{real*4} numbers).

When core SD-amplitudes are found, we find one-electron valence amplitudes
from the system of equations from Fig.~\ref{fig3} and~\ref{fig4}. This
allows us to form $H_{\rm eff}$ for one electron above the core and find
the spectrum of the three-electron system. These results are compared to
the three-electron full CI in \Tref{tab1}.

Finally, we calculate two-electron valence SD-amplitudes and form
two-electron effective Hamiltonian. Two-electron full CI with $H_{\rm eff}$
is compared with four-electron full CI in \Tref{tab2}.

\section{Discussion}

SD-method for the system with valence electrons is not exact even for the
two-electron core~\cite{BJLS89}. Therefore we can not expect exact
agreement between SD and full CI results. We still expect higher
accuracy for the CI+SD method than for CI with the second order $H_{\rm
eff}$. Tables~\ref{tab1} and~\ref{tab2} show that this is the case. For
the one-electron case both the maximum and the average error decreases by
the factor of 3 to 4. For the two-electron case the error decreases even
stronger.

Moreover, if the third order corrections are added to the one-electron
SD-amplitudes, as was suggested in~\cite{SJD99}, the accuracy rises by
almost another order of magnitude. If we include one-electron third order
corrections to the two-electron effective Hamiltonian, the error in
comparison to the SD-approximation even grows. The total number of the
third order two-electron diagrams is very large and at present we are able
to include them only partly. It is seen from the Table~\ref{tab2} that this
improves the accuracy a bit in comparison to the SD-approximation. However,
we see that there are no dominant third order diagrams, which are not
included in the SD-approximation. In fact there are many contributions of
the same order of magnitude and of different signs, so that they strongly
cancel each other. It is possible that complete third order correction
would give even better accuracy, but any partial account of the third order
may be dangerous.

We conclude that in the many-electron case there is no point in including
only one-electron third order corrections to the effective Hamiltonian or
any other subset of the complete set of the third order terms. That
probably means that for a more complicated atom it may be too difficult to
improve SD-results by including the third order corrections. Note that for
more than two valence electrons there will be also three-electron terms in
the effective Hamiltonian.

We see that for the simple model system considered here the
SD-approximation for the effective Hamiltonian appears to be few times more
accurate than the second order MBPT. It is still not obvious that this will
hold for the more complicated systems. As we saw above, the two-electron
core is a special case for the SD method as the latter allows to obtain the
core energy exactly here. The advantage of the model considered here is
that we can compare results with the full-electron calculation, which is
impossible in the general case. The computational expenses for a full
scale atomic calculations within CI+SD method are significantly higher
that in CI+MBPT(II), but are still lower than in CI+MBPT(III).

\smallskip

The author is grateful to E.\,Eliav, T.\,Isaev, W.\,Johnson, N.\,Mosyagin,
S.\,Porsev, and A.\,Titov for valuable discussions. This work was supported
by RFBR, grant No 02-02-16387.


\newpage 
\begin{table}
\caption{Comparison of different one-electron approximations with
three-electron full CI. Three-electron transition frequencies and the
errors for one-electron calculations are given in cm$^{-1}$. One-electron
calculations include Dirac-Fock (DF), Dirac-Fock with the second order
self-energy correction (MBPT), SD-approximation, and SD with third order
corrections (SD+III). Averaged absolute value of the errors and maximum
errors are given at the bottom.}

\label{tab1}
\begin{tabular}{lrrrrr}
\hline
Level
& \multicolumn{1}{c}{Full CI}
& \multicolumn{4}{c}{One-electron approximations}\\
&& \multicolumn{1}{c}{DF}
& \multicolumn{1}{c}{MBPT}
& \multicolumn{1}{c}{SD}
& \multicolumn{1}{c}{SD+III}\\
\hline
$2s_{1/2}$ &       0.0 &    0.0 &    0.0 &    0.0 &   0.0  \\
$2p_{1/2}$ &   64870.1 &  329.9 &   46.1 &   16.0 &   2.2  \\
$2p_{3/2}$ &   64999.6 &  328.6 &   45.2 &   15.1 &   1.5  \\
$3s_{1/2}$ &  302411.3 & -256.1 &  -49.5 &  -11.1 &  -1.2  \\
$3p_{1/2}$ &  319727.1 & -189.4 &  -40.8 &   -8.1 &  -1.1  \\
$3p_{3/2}$ &  319765.1 & -189.7 &  -40.9 &   -8.4 &  -1.4  \\
$3d_{3/2}$ &  324342.0 & -284.9 &  -61.7 &  -13.9 &  -2.2  \\
$3d_{5/2}$ &  324351.2 & -284.9 &  -61.7 &  -13.8 &  -2.2  \\
\hline
av. err.   &           &  266   &    50  &     12 &   1.7  \\
max. err.  &           &  330   &    50  &     16 &   2.2  \\
\hline
\end{tabular}
\end{table}

\begin{table}
\caption{Comparison of two-electron approximations with four-electron full CI.
Two-electron full CI is made for initial Hamiltonian and for effective
Hamiltonians calculated within second order MBPT, SD, and SD with
third order corrections (SD+III). In the latter case either one-electron or
one-electron and partly two-electron terms were included.}

\label{tab2}
\begin{tabular}{lrrrrrr}
\hline
Level
& \multicolumn{1}{c}{Full CI}
& \multicolumn{5}{c}{Two-electron approximations}\\
&& \multicolumn{1}{c}{CI}
& \multicolumn{1}{c}{MBPT}
& \multicolumn{1}{c}{SD}
& \multicolumn{2}{c}{SD+III}\\
\hline
$^1S_0(2s^2)$&      0.0 &      0.0 &   0.0 &   0.0 &   0.0 &  0.0 \\
$^3P_0(2s2p)$&  52535.9 &    222.0 &  31.6 &   5.0 &  -4.9 &  1.9 \\
$^3P_1(2s2p)$&  52569.7 &    221.9 &  31.8 &   5.2 &  -4.7 &  2.1 \\
$^3P_2(2s2p)$&  52637.5 &    221.7 &  31.8 &   5.7 &  -4.2 &  2.6 \\
$^1P_1(2s2p)$& 104969.2 &    865.4 &  80.3 & -19.9 & -30.1 &  9.7 \\
$^3P_0(2p^2)$& 138303.7 &   1016.9 & 111.0 &  -1.8 & -22.0 &  0.4 \\
$^3P_1(2p^2)$& 138337.0 &   1016.3 & 110.7 &  -2.2 & -22.4 & -0.2 \\
$^3P_2(2p^2)$& 138402.9 &   1015.3 & 110.4 &  -2.4 & -22.6 & -0.2 \\
$^1D_2(2p^2)$& 147954.8 &    698.6 &  70.1 &  -3.5 & -21.5 &  9.7 \\
$^1S_0(2p^2)$& 186561.7 &    845.5 &  63.7 & -14.7 & -45.4 & 13.7 \\
$^3S_1(2s3s)$& 239752.9 &     79.5 &  11.7 &  13.7 &  22.0 &  8.4 \\
$^1S_0(2s3s)$& 251971.5 &     59.8 &   0.9 &   2.0 &   9.7 &  1.6 \\
\hline
av. err.     &      0   &    569   &  60   &   7   &  19   &  5   \\
max. err.    &      0   &   1017   & 111   &  20   &  45   & 14   \\
\hline
\end{tabular}
\end{table}

\newpage 

\end{document}